\documentclass[
 aps,%
 prl,%
 tightenlines,
 preprint,preprintnumbers,%
 groupedaddress,superscriptaddress,
 showpacs,%
 amssymb, amsmath%
]{revtex4}
\usepackage{graphicx}
\usepackage{xspace}
\usepackage{dcolumn}
\usepackage{epsfig}

\newcommand{\Bz}{B^0}
\newcommand{\Bzbar}{\overline{B}{}^0}
\newcommand{\rto}{\rightarrow}
\newcommand{\ThisLum}{140~{\rm fb}^{-1}}
\newcommand{\Bzrp}{\Bz\rto\rho^\pm\pi^\mp}
\newcommand{\Bzbarrp}{\Bzbar\rto\rho^\mp\pi^\pm}

\newcommand{\ArpCP}{{\cal A}^{\rho\pi}_{CP}}
\newcommand{\Arpvalue}{-0.16}
\newcommand{\Arpstat}{\pm 0.10}
\newcommand{\Arpsyst}{\pm 0.02}

\newcommand{\CrpCP}{C_{\rho\pi}}
\newcommand{\Crpvalue}{0.25}
\newcommand{\Crpstat}{\pm 0.17}
\newcommand{\Crpsyst}{^{+0.02}_{-0.06}}
\newcommand{\Crpallerr}{\pm 0.17^{+0.02}_{-0.06}}

\newcommand{\dCrpCP}{\Delta\CrpCP}
\newcommand{\dCrpvalue}{0.38}
\newcommand{\dCrpstat}{\pm 0.18}
\newcommand{\dCrpsyst}{^{+0.02}_{-0.04}}
\newcommand{\dCrpallerr}{\pm 0.18^{+0.02}_{-0.04}}

\newcommand{\SrpCP}{S_{\rho\pi}}
\newcommand{\Srpvalue}{-0.28}
\newcommand{\Srpstat}{\pm 0.23}
\newcommand{\Srpsyst}{^{+0.10}_{-0.08}}
\newcommand{\Srpallerr}{\pm 0.23^{+0.10}_{-0.08}}

\newcommand{\dSrpCP}{\Delta\SrpCP}
\newcommand{\dSrpvalue}{-0.30}
\newcommand{\dSrpstat}{\pm 0.24}
\newcommand{\dSrpsyst}{\pm 0.09}
\newcommand{\dSrpallerr}{\pm 0.24\pm 0.09}

\newcommand{\dt}{\Delta t}

\newcommand{\Tbz}{\tau_{B^0}}
\newcommand{\dm}{\Delta m_d}
\newcommand{\ArKCP}{{\cal A}^{\rho K}}

\newcommand{\Srp}{\SrpCP}
\newcommand{\Crp}{\CrpCP}
\newcommand{\Apm}{{\cal A}^{+-}_{\rho \pi}}
\newcommand{\Apmvalue}{-0.02}
\newcommand{\Apmstat}{\pm 0.16}
\newcommand{\Apmsyst}{^{+0.05}_{-0.02}}
\newcommand{\Apmallerr}{\pm 0.16^{+0.05}_{-0.02}}
\newcommand{\Amp}{{\cal A}^{-+}_{\rho \pi}}
\newcommand{\Ampvalue}{-0.53}
\newcommand{\Ampstat}{\pm 0.29}
\newcommand{\Ampsyst}{^{+0.09}_{-0.04}}
\newcommand{\Ampallerr}{\pm 0.29^{+0.09}_{-0.04}}
\newcommand{\DCRP}{\Delta{C}_{\rp}}

\newcommand{\rp}{\rho\pi}

\newcommand{\BBbar}{B\overline{B}}

\newcommand{\NBBBelle}{152\times 10^6}

\newcommand{\qq}{q\bar{q}}

\begin{document}

\vspace*{-3\baselineskip}
\epsfysize3cm
\epsfbox{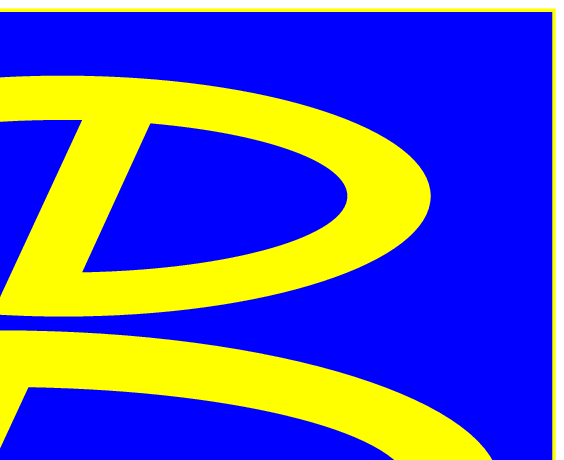}


\preprint{\vbox{\hbox{}\hbox{}
 		\hbox{Belle preprint 2004--21}
                \hbox{KEK preprint 2004--34}
		\hbox{hep-ex/0408003}
                }}

\title{\large\boldmath
       Study of $\Bzrp$ Time-dependent CP Violation 
       at Belle}

\affiliation{Budker Institute of Nuclear Physics, Novosibirsk}
\affiliation{Chonnam National University, Kwangju}
\affiliation{University of Cincinnati, Cincinnati, Ohio 45221}
\affiliation{University of Frankfurt, Frankfurt}
\affiliation{Gyeongsang National University, Chinju}
\affiliation{University of Hawaii, Honolulu, Hawaii 96822}
\affiliation{High Energy Accelerator Research Organization (KEK), Tsukuba}
\affiliation{Hiroshima Institute of Technology, Hiroshima}
\affiliation{Institute of High Energy Physics, Chinese Academy of Sciences, Beijing}
\affiliation{Institute of High Energy Physics, Vienna}
\affiliation{Institute for Theoretical and Experimental Physics, Moscow}
\affiliation{J. Stefan Institute, Ljubljana}
\affiliation{Kanagawa University, Yokohama}
\affiliation{Korea University, Seoul}
\affiliation{Kyungpook National University, Taegu}
\affiliation{Swiss Federal Institute of Technology of Lausanne, EPFL, Lausanne}
\affiliation{University of Ljubljana, Ljubljana}
\affiliation{University of Maribor, Maribor}
\affiliation{University of Melbourne, Victoria}
\affiliation{Nagoya University, Nagoya}
\affiliation{Nara Women's University, Nara}
\affiliation{National Central University, Chung-li}
\affiliation{National Kaohsiung Normal University, Kaohsiung}
\affiliation{National United University, Miao Li}
\affiliation{Department of Physics, National Taiwan University, Taipei}
\affiliation{H. Niewodniczanski Institute of Nuclear Physics, Krakow}
\affiliation{Nihon Dental College, Niigata}
\affiliation{Niigata University, Niigata}
\affiliation{Osaka City University, Osaka}
\affiliation{Osaka University, Osaka}
\affiliation{Panjab University, Chandigarh}
\affiliation{Peking University, Beijing}
\affiliation{Princeton University, Princeton, New Jersey 08545}
\affiliation{University of Science and Technology of China, Hefei}
\affiliation{Seoul National University, Seoul}
\affiliation{Sungkyunkwan University, Suwon}
\affiliation{University of Sydney, Sydney NSW}
\affiliation{Tata Institute of Fundamental Research, Bombay}
\affiliation{Toho University, Funabashi}
\affiliation{Tohoku Gakuin University, Tagajo}
\affiliation{Tohoku University, Sendai}
\affiliation{Department of Physics, University of Tokyo, Tokyo}
\affiliation{Tokyo Institute of Technology, Tokyo}
\affiliation{Tokyo Metropolitan University, Tokyo}
\affiliation{Tokyo University of Agriculture and Technology, Tokyo}
\affiliation{University of Tsukuba, Tsukuba}
\affiliation{Virginia Polytechnic Institute and State University, Blacksburg, Virginia 24061}
\affiliation{Yonsei University, Seoul}
  \author{C.~C.~Wang}\affiliation{Department of Physics, National Taiwan University, Taipei} 
  \author{K.~Abe}\affiliation{High Energy Accelerator Research Organization (KEK), Tsukuba} 
  \author{K.~Abe}\affiliation{Tohoku Gakuin University, Tagajo} 
  \author{T.~Abe}\affiliation{High Energy Accelerator Research Organization (KEK), Tsukuba} 
  \author{I.~Adachi}\affiliation{High Energy Accelerator Research Organization (KEK), Tsukuba} 
  \author{H.~Aihara}\affiliation{Department of Physics, University of Tokyo, Tokyo} 
  \author{Y.~Asano}\affiliation{University of Tsukuba, Tsukuba} 
  \author{T.~Aushev}\affiliation{Institute for Theoretical and Experimental Physics, Moscow} 
  \author{A.~M.~Bakich}\affiliation{University of Sydney, Sydney NSW} 
  \author{Y.~Ban}\affiliation{Peking University, Beijing} 
% ...
\collaboration{The Belle Collaboration}

\begin{abstract}
We present a time-dependent analysis of $\textit{CP}$ violation in
$\Bzrp$ decays based on a $\ThisLum$ data sample collected at the
$\Upsilon(4S)$ resonance with the Belle detector at the KEKB
asymmetric-energy $e^+e^-$ collider. We fully reconstruct one
neutral $\textit{B}$ meson in the $\rho^\pm\pi^\mp$ final state
and identify the flavor of the accompanying $\textit{B}$ meson
from its decay products. We obtain the charge asymmetry
$\ArpCP = \Arpvalue {\Arpstat}$(stat)${\Arpsyst}$(syst).
An unbinned maximum likelihood fit to the proper-time distributions yields
$\CrpCP = \Crpvalue {\Crpstat}$(stat)${\Crpsyst}$(syst),
$\dCrpCP = \dCrpvalue {\dCrpstat}$(stat)$\dCrpsyst$(syst),
$\SrpCP = \Srpvalue {\Srpstat}$(stat)${\Srpsyst}$(syst), and
$\dSrpCP = \dSrpvalue {\dSrpstat}$(stat)$
\dSrpsyst$(syst). The direct $CP$ violation parameters for 
$B\to \rho^+\pi^-$ and $B\to \rho^-\pi^+$ decays are 
$\Apm = \Apmvalue {\Apmstat}$(stat)${\Apmsyst}$(syst) and
$\Amp = \Ampvalue {\Ampstat}$(stat)${\Ampsyst}$(syst).
\end{abstract}


\maketitle

In the Standard Model (SM) of elementary particles, $CP$ violation arises 
from the Kobayashi-Maskawa (KM) phase~\cite{bib:KM} in the 
weak-interaction quark-mixing matrix. Recently, the
Belle~\cite{bib:belle_pipi} and BaBar
collaborations~\cite{bib:babar_pipi} reported results on $CP$ violation via
$b \to u\bar{u}d$ transitions in $B\to \pi^+\pi^-$ decays,
which are related to the $CP$ violation parameter $\phi_2$. Here we present
a study of $B\to\rho^\pm\pi^\mp$, which is another $\phi_2$ related decay.
Since $B\to\rho^\pm\pi^\mp$ is not a $CP$ eigenstate decay, four decay
modes with different charge and flavor combinations in the neutral $B$
system must be considered. 

In the decay chain 
$\Upsilon(4S)\to B^0\overline{B}{}^0 \to (\rho^\pm \pi^\mp) f_{\rm tag}$, 
one of the $B$ mesons decays at time $t_{\rho\pi}$ to $\rho^\pm\pi^\mp$ 
and the other meson decays at time $t_{\rm tag}$ to a final state 
$f_{\rm tag}$ that distinguishes between $B^0$ and $\overline{B}{}^0$. 
The decay rate for $B^0(\overline{B}{}^0)\to \rho^\pm\pi^\mp$ has a 
time dependence given by
\begin{eqnarray}
  \label{eq:pdf}
  {\cal P}^{\rho^{\pm}\pi^{\mp}}_{q}(\dt)
  & =& (1\pm\ArpCP)\frac{e^{-|\Delta t|/{\Tbz}}}{8{\Tbz}}\\\nonumber
  &&\times\{1+q{\cdot}
  [(\Srp\pm\Delta\Srp)\sin(\dm\dt) \\\nonumber
  & &-(\Crp\pm\Delta\Crp)\cos(\dm\dt)]\},
\end{eqnarray}
where $\Tbz$ is the $B^0$ lifetime, $\dm$ is the mass difference between 
the two $B^0$ mass eigenstates, $\Delta t = t_{\rho\pi} -  t_{\rm tag}$, 
and the $b$-flavor charge $q$ = +1($-1$) when the tagging $B$ meson is a 
$B^0(\overline{B}{}^0)$. The time and flavor integrated charge asymmetry 
$\ArpCP$ is defined as 
\begin{equation}
  \ArpCP =\frac{N(\rho^+\pi^-)-N(\rho^-\pi^+)}{N(\rho^+\pi^-)+N(\rho^-\pi^+)},
\end{equation}
where $N(\rho^+\pi^-)$ and $N(\rho^-\pi^+)$ are the sum of the yields for
$\Bz$ and $\Bzbar$ decays to $\rho^+\pi^-$ and $\rho^-\pi^+$, respectively. 
The mixing-induced $CP$ violation parameter $\Srp$ is related to $\phi_2$
and $\Crp$ is the flavor-dependent direct $CP$ violation parameter.
The asymmetry between the decay rates, 
$\Gamma(B^0\to \rho^+\pi^-)+\Gamma(\overline{B}{}^0\to\rho^-\pi^+)$ and
$\Gamma(B^0\to \rho^-\pi^+)+\Gamma(\overline{B}{}^0\to\rho^+\pi^-)$, is
described by $\Delta\Crp$, while the strong phase difference between the
amplitudes contributing to $B^0\to\rho\pi$ decays is described by $\Delta\Srp$.
These parameters are related to $\phi_2$ as $\Srp\pm\Delta\Srp$ =
$\sqrt{1-(\Crp\pm\Delta\Crp)^2}\sin(2 \phi  ^\pm_{2\rm eff}\pm\delta)$,
where 2$\phi ^\pm_{2 \rm eff}$ =
arg[$(q/p)({\bar{A}^\pm_{\rho\pi}}/{A^\mp_{\rho\pi}})$] and
$\delta$ = arg[$A^-_{\rho\pi}/A^+_{\rho\pi}$];
arg[$q/p$] is the $\Bz$-$\Bzbar$ mixing phase.
The terms $A^+_{\rho\pi}(\bar{A}^+_{\rho\pi})$ and
$A^-_{\rho\pi}(\bar{A}^-_{\rho\pi})$ denote the transition amplitudes for 
the processes $\Bz(\Bzbar)\rto\rho^+\pi^-$ and $\Bz(\Bzbar)\rto\rho^-\pi^+$,
respectively. The angles $\phi ^\pm_{2 \rm eff}$ are equal to $\phi _2$
if there is no penguin contribution.
The effect of direct $CP$ violation can also be expressed in
terms of another set of parameters, $\Apm$ and $\Amp$~\cite{bib:phi2_su3}:
\begin{eqnarray}
{\cal A}_{\rp}^{\pm\mp} &=& \frac{N(\Bzbarrp)-N(\Bzrp)}%
      {N(\Bzbarrp)+N(\Bzrp)} \\
&=& \mp\frac{\ArpCP\pm\Crp\pm\ArpCP\cdot{\DCRP}}%
            {1\pm{\DCRP}\pm\ArpCP\cdot\Crp}\nonumber
\end{eqnarray}

The strategy of this analysis is to reconstruct final states in quasi-two-body
decays $\Bz\to(\pi^\pm\pi^0)\pi^\mp$, which correspond to distinct bands in
the $\pi^+\pi^-\pi^0$ Dalitz plot. We exclude the interference region where 
the $\rho$ charge is ambiguous, and neglect possible residual interference 
effects.

The results for this analysis are based on $\ThisLum$ of integrated luminosity,
which corresponds to $\NBBBelle$ produced $\BBbar$ pairs. The data were 
collected with the Belle detector at the KEKB asymmetric-energy $e^+e^-$ 
collider~\cite{bib:KEKB}, which collides 8.0 GeV $e^-$ and 3.5 GeV $e^+$ 
beams. The $\Upsilon(4S)$ is produced with a Lorentz boost of 
$\beta\gamma=0.425$ nearly along the electron beamline. 
Since the $\Bz$ and $\Bzbar$ mesons are approximately at rest in the 
$\Upsilon (4S)$ center-of-mass system (CM), $\dt$ can be determined from 
$\Delta z$, the displacement in $z$ between the $\rho^\pm\pi^\mp$ and 
$f_{\rm tag}$ decay vertices:
$\dt \simeq (z_{\rho^\pm\pi^\mp}-z_{\rm tag})/\beta\gamma c$. 
The $z$ axis is anti-parallel to the positron beam.

The Belle detector~\cite{bib:Belle} is a large-solid-angle general purpose 
spectrometer that consists of a silicon vertex detector (SVD), a central drift 
chamber (CDC), an array of aerogel threshold \v{C}erenkov counters (ACC), 
time-of-flight scintillation counters (TOF), and an electromagnetic calorimeter 
comprised of CsI(Tl) crystals located inside a superconducting solenoid coil
that provides a 1.5~T magnetic field. An iron flux return located outside of
the coil is instrumented to detect $K_L^0$ mesons and identify muons.

To reconstruct $\Bzrp$ candidates, we combine pairs of oppositely 
charged tracks with $\pi^0$ candidates. Each charged track is required to 
have transverse momenta greater than $100~\textrm{MeV}/c$ in the laboratory frame.  
Charged tracks are identified as pions by combining information from the 
ACC, CDC and TOF. Electron-like tracks are rejected.
The $\gamma$ energies for $\pi^0$ candidates are required to be 
greater than 50 MeV if the photon is detected in the barrel ECL
($32^\circ < \theta < 129^\circ$);
otherwise, the energy is required to be larger than 100 MeV,
where $\theta$ denotes the polar angle with respect to the $z$-axis.
The $\pi^0$ candidates are selected from $\gamma \gamma$ pairs with
invariant masses in the range 
$0.118~\textrm{GeV}/c^2 <M_{\gamma\gamma}<0.150~\textrm{GeV}/c^2$,
and momentum larger than 200 MeV/$c$ in the laboratory frame. 
In addition, we require
$|\cos \theta^{\pi^0}_{\rm dec}|<0.95$, 
where $\theta^{\pi^0}_{\rm dec}$ is defined as the angle between
the photon flight direction and the boost direction from the
laboratory system in the $\pi^0$ rest frame,
and we require the $\chi^2$ of the $\pi^0$
mass-constrained fit to be less than 50.

$B$ meson candidates are reconstructed using the beam-energy constrained mass
$M_{\rm bc}\equiv\sqrt{{E^2_{\rm beam}} - {P^2_B}}$
and the energy difference
$\Delta E \equiv E_{B} - E_{\rm beam}$. The variables $E_B$ and $P_B$ are the 
reconstructed energy and
momentum of the $B$ candidate in the CM frame, and $E_{\rm beam}$ is the CM 
beam energy. The $B$ candidates in the region with $M_{\rm bc}>5.2$ GeV/$c^2$ 
and $-0.3~{\rm GeV} < \Delta E <0.2$~GeV are selected.
The signal region is defined as 
$M_{\rm bc}>5.27$ GeV/$c^2$ and $-0.10 {\rm ~GeV} < \Delta E <0.08$ GeV.
The $B\to\rho^\pm\pi^\mp$ candidates are formed from 3-body
$B\to \pi^+\pi^-\pi^0$ decays with a $\pi^\pm\pi^0$ invariant mass in the range
0.57 GeV$/c^2 <M_{\pi^\pm\pi^0}<0.97$ GeV/$c^2$ and $\rho$ helicity 
$|\cos\theta^\rho_{\rm hel}|>0.5$, 
where $\theta^\rho_{\rm hel}$ is defined as the angle between the charged pion
direction and the $B^0$ direction in the $\rho$ rest frame.
To avoid the region where the $\rho^+\pi^-$ and $\rho^-\pi^+$ contributions
interfere, we exclude candidates with both 
$M_{\pi^+\pi^0}$ and $M_{\pi^-\pi^0}$ smaller than 1.22 GeV/$c^2$.
Candidates with $M_{\pi^+\pi^-}<0.97$ GeV/$c^2$ are removed to avoid
the region where the $\rho^+\pi^-$ or $\rho^- \pi^+$ bands
overlap with $\rho^0\pi^0$. 

To suppress the dominant $e^+e^-\to q\bar{q}$ continuum background 
($q =  u,d,s,c$), we form the likelihood ratio 
$\mathcal{R}=\mathcal{L}_{\rm s}/(\mathcal{L}_{\rm s}+\mathcal{L}_{\rm bkg})$, 
where $\mathcal{L}_{\rm s}$ and $\mathcal{L}_{\rm bkg}$ are
likelihood functions for signal and continuum events, respectively.
We use a Fisher discriminant based on five modified Fox-Wolfram 
moments~\cite{bib:sfw}, and the CM flight direction of the $B$ ($\theta_B$)
with respect to the $z$-axis to form the likelihood function. 
The signal likelihood $\mathcal{L}_{\rm s}$ is determined from a
GEANT-based Monte Carlo (MC), and $\mathcal{L}_{\rm bkg}$ is based on 
$M_{\rm bc}$ sideband data, 
$M_{\rm bc}<$ 5.26 GeV/$c^2$. 
The continuum background is reduced by requiring $\mathcal{R}$
to be greater than 0.8.
If there is more than one candidate in an event, 
we select the candidate with the smallest
sum of the $\chi^2$ for the $\pi^+\pi^-$ vertex fit and 
the $\pi^0$ mass-constrained fit.

The flavor of the accompanying $B$ meson is identified from the decay products
not associated with the reconstructed $\Bzrp$ decay. We use the same method 
as used for the Belle $\sin2\phi_1$ measurement
\cite{bib:belle_phi1,bib:fbtag}. Two parameters $q$ and $r$ are used to 
describe the flavor tagging information. The parameter $q$ is defined in 
Eq.~\ref{eq:pdf}, and the parameter $r$ is a MC-determined quality factor 
that ranges from $r=0$ for no flavor discrimination to $r=1$ for unambiguous 
flavor assignment. It is used only to sort data into six $r$ intervals.
The wrong tag fractions for $B^0$
and $\overline{B}{}^0$ are obtained from $B\to D^*l\nu$, $D^*\pi$,
$D^*\rho$, and $D\pi$ data for the six $r$ intervals.

\begin{figure}
\begin{center}
\includegraphics[width=0.95\columnwidth]{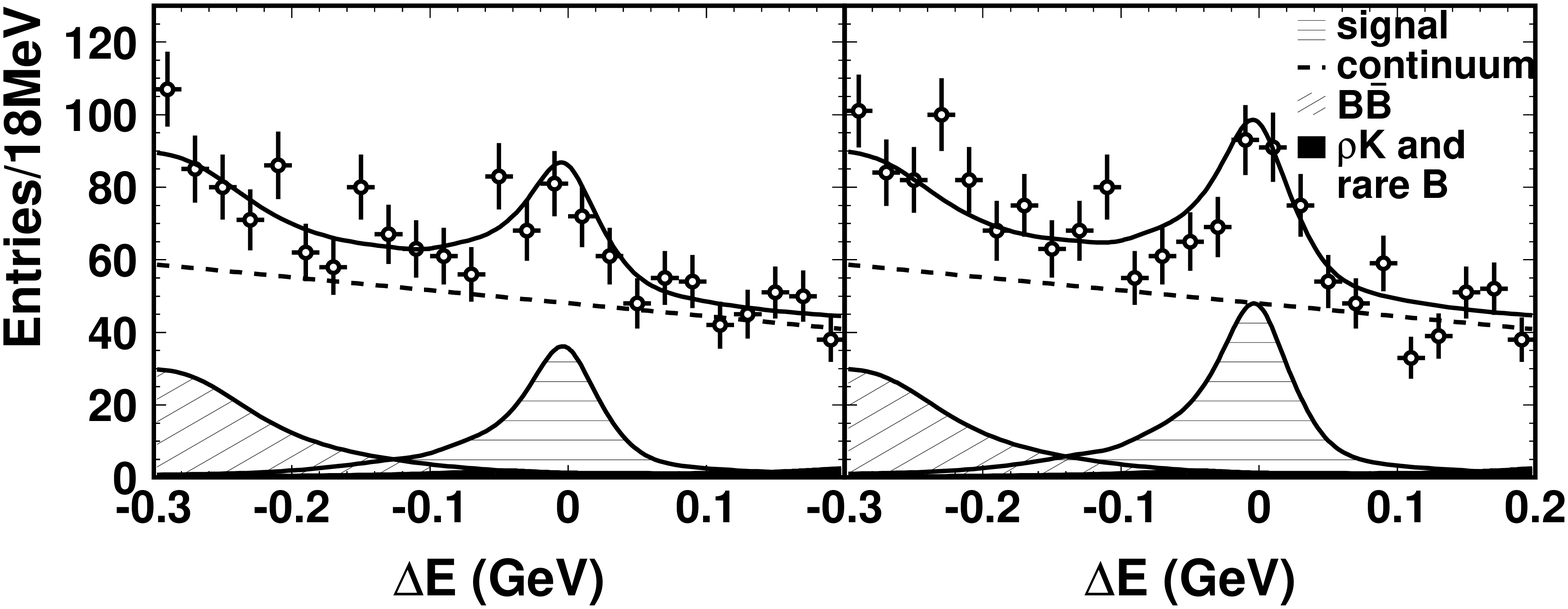}
\includegraphics[width=0.95\columnwidth]{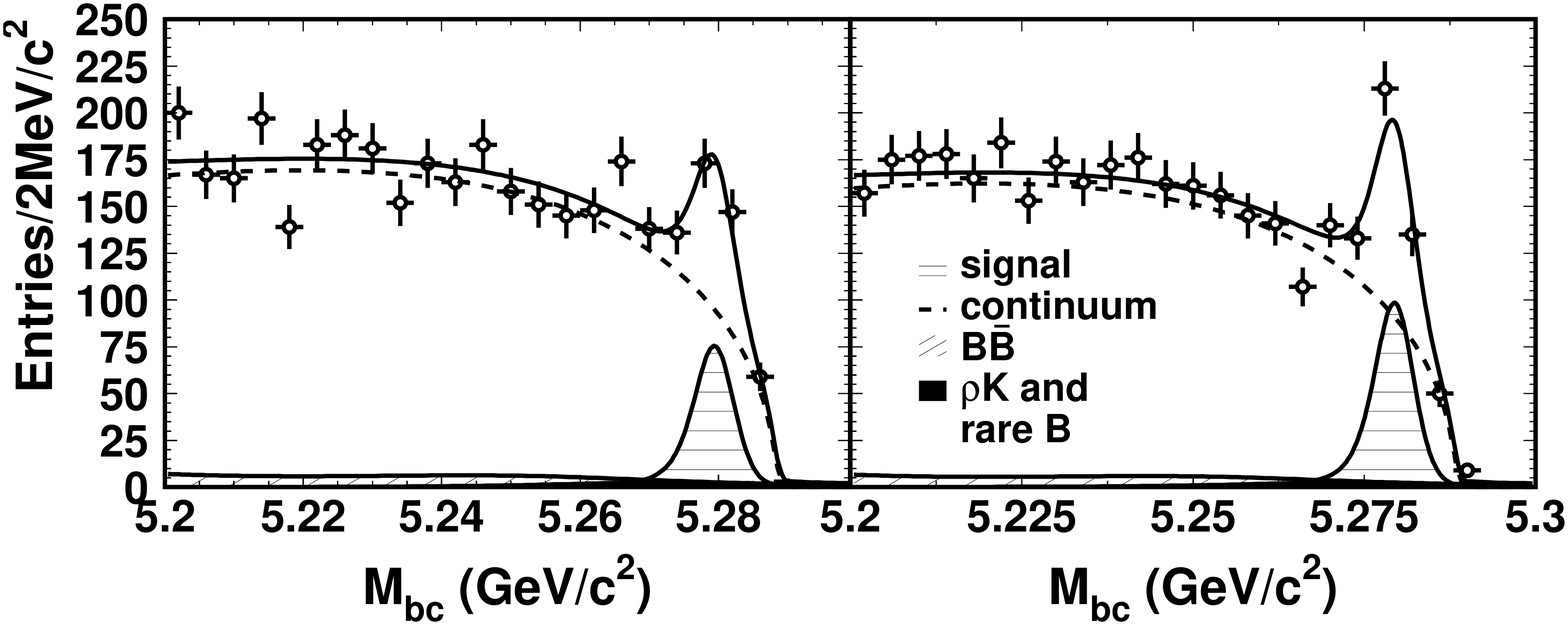}
\end{center}
\caption{$\Delta E$ (top) and $M_{\rm bc}$ (bottom) projections for the 
result of the 2-D unbinned likelihood fit.
The plots on the left are the results for the $\rho^+\pi^-$ candidates, 
while those on the right show the results for the $\rho^-\pi^+$ candidates.
}
\label{fig:demb}
\end{figure}

The vertex reconstruction algorithm is the same as that used for the 
$\sin2\phi_1$ analysis~\cite{bib:belle_phi1}. The vertex positions for 
$\rho^\pm \pi^\mp$ and $f_{\rm tag}$ decays are
reconstructed from charged tracks with associated SVD hits and an 
interaction point constraint. The vertex for $f_{\rm tag}$ is determined from
all well-reconstructed tracks excluding the tracks from the 
$\Bzrp$ decay and $K^0_S$ candidates.

Figure~\ref{fig:demb} shows the $\Delta E$ ($M_{\rm bc}$) distribution in 
the $M_{\rm bc}$ ($\Delta E$) signal region for $\Bzrp$ candidates after 
flavor tagging and vertex reconstruction.  
The $\rho^\pm\pi^\mp$ signal yields are extracted from an unbinned
maximum-likelihood fit to the two-dimensional ($M_{\rm bc}$,$\Delta E$)
distribution. The backgrounds are categorized as continuum $q\bar{q}$,
$b\to c$ transitions ($\BBbar$), $B\to \rho K$, and 
rare charmless decays other than $B\to \rho K$ (rare $B$).
The distributions for $\rho\pi$, $\BBbar$, $\rho K$, and
rare $B$ events are obtained from MC.

\begin{figure}
\begin{center}
\includegraphics[width=\columnwidth]{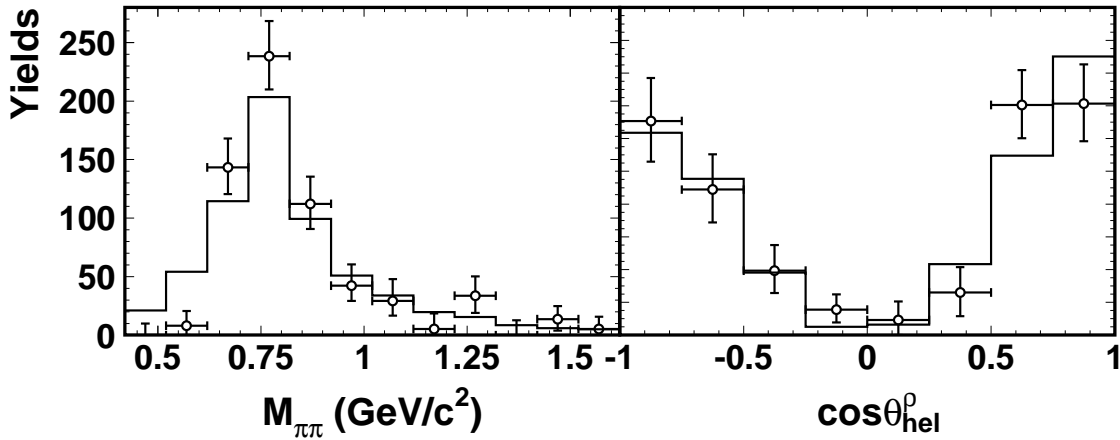}
\end{center}
\caption{
Signal yields as functions of (left) $M_{\pi\pi}$ and (right) 
$\cos\theta^{\rho}_{\rm hel}$ in data. The histograms show
the results of $B\to\rho^{\pm}\pi^{\mp}$ MC simulation with areas 
normalized to the total signal yield.
}
\label{fig:rho}
\end{figure}

The $\rho\pi$ signal PDF contains two components: signal events reconstructed 
with the correct charge (${P}_{\rho\pi}$) and those with incorrect charge
(${P}^{\rm wc}_{\rho\pi}$). The fraction of events with incorrect 
charge in the signal region due to combinations that include a random $\pi^0$
is estimated to be 2.7\% from MC and is fixed in the fit.
The signal PDF shape is modeled by a smoothed histogram.
The $\Delta E$ distributions for $B\to \rho\pi$ signal are parameterized
separately for $\pi^0$ momentum below and above 1.2 GeV/$c$
in the laboratory frame.
The $\Delta E$ widths for $\rho\pi$ and $\rho K$ are calibrated from
$D^{*0}\to D^0[K^-\pi^+]\pi^0$ data.
The $B^+\to D^0[K^-\pi^+\pi^0] \pi^+$ mode is used to calibrate the 
$\Delta E$ and $M_{\rm bc}$ peak positions. 
The $M_{\rm bc}$ and $\Delta E$ distributions for the continuum 
$q\bar{q}$ are parametrized 
by an ARGUS background function~\cite{bib:argus} and a linear function, 
respectively.
The contributions from $B\to \rho K$ 
(with ${\cal B} = (9.0\pm1.6) \times 10^{-6}$~\cite{bib:rhok_wa}) 
and from rare $B$ decays are fixed in the fit, while the yields  
for $B\to \rho \pi$ signal, 
$B\overline {B}$ and continuum backgrounds, and the shape parameters for 
continuum are floated. We obtain $483\pm46$
$B\to \rho^\pm\pi^\mp$ events, and obtain a time and flavor integrated 
charge asymmetry $\ArpCP=\Arpvalue \Arpstat$(stat). The estimated yields
for $B\to\rho \pi$, $B\to \rho K$, $\qq$, $\BBbar$ and rare $B$ in the 
signal region are 328.7, 11.2, 833.0, 23.3 and 18.8, respectively.
We remove the requirements on $M_{\pi^\pm\pi^0}$ and 
$\cos\theta^{\rho}_{\rm hel}$ and examine these distributions to verify 
that the signals reconstructed as $B\to \pi^+\pi^-\pi^0$ are from the 
two-body decay $B\to \rho\pi$. Figure~\ref{fig:rho} shows the signal yields 
in bins of $M_{\pi^\pm\pi^0}$ and $\cos \theta^{\rho}_{\rm hel}$ for data.

The $CP$ violation parameters are obtained from an unbinned 
maximum-likelihood fit to the observed proper-time distribution for the
$B\to \rho\pi$ candidates in the  ($M_{\rm bc}$,$\Delta E$) signal region.
The likelihood function describing the proper-time distribution is
\begin{eqnarray}
{\cal L}  &=&\prod^{N}_{i=1}\{f_{\rho\pi}{P}_{\rho\pi}(\dt_i)
	   + f_{\rho\pi}^{\rm wc}{P}^{\rm wc}_{\rho\pi}(\dt_i)\\
          &+&f_{\rho{K}}         {P}_{\rho K}(\dt_i)
           + f_{\qq}             {P}_{\qq}(\dt_i)\nonumber\\
          &+&f_{\BBbar}          {P}_{\BBbar}(\dt_i)
           + f_{\textrm{rare}B}  {P}_{\textrm{rare}B}(\dt_i)\},\nonumber
\end{eqnarray}
where the weighting functions $f_m$ ($m$ = $\rho\pi$, $\rho{K}$, $\qq$, 
$\BBbar$, and rare $B$) are determined on an event-by-event basis as
functions of $M_{\rm bc}$ and $\Delta E$ 
for each flavor tagging $r$ interval and $\pi^0$ momentum range in the 
laboratory system. 
The time-dependent probability density functions
($\Delta t$ PDFs)
${P}_{\rm \rho\pi}(\dt_i)$ for $B\to \rho\pi$
 and ${P}_{\rho{K}}(\dt_i)$ for $B\to \rho K$
are obtained from the true PDFs convolved with the $\Delta t$ 
resolution function used in the
$\sin2\phi_1$ measurement~\cite{bib:belle_phi1}. 
The true PDF for $B\to \rho \pi$ is given by Eq.1 modified to incorporate the
effect of incorrect flavor tagging.  
The PDF for $B\to \rho\pi$ signal reconstructed with incorrect charge,
${P}_{\rm \rho^\pm\pi^\mp}^{\rm wc}(\dt_i)$, is given by 
${P}_{\rm \rho^\mp\pi^\pm}(\dt_i)$.
For $B\to \rho K$, $C = S = \Delta S = 0$,
$\Delta C = -1$, and $\ArKCP = 0$ is assumed.
The resolution function consists
of the detector resolution, the shift in vertex position due to secondary 
tracks originating from charmed particle decays, and smearing due to the 
approximation $\dt \simeq (z_{\rho^\pm\pi^\mp}-z_{\rm tag})/\beta\gamma c$.
The $\Delta t$ PDFs for other backgrounds are all parameterized as
${\cal P}_j = (1 - f_j) \delta(\Delta t - \mu^j_\delta) + 
   f_j {\rm exp}(- \frac{|\Delta t - \mu^j_\tau|}{\tau_j}) $
convolved with $R_j$ ($j$ = $q\bar q$, $B\overline{B}$ and rare B),
where $f_j$ is the fraction of the background with effective lifetime
$\tau_j$.
The resolution-like function $R_j$ for background is given by two Gaussians.
The parameters of the $\Delta t$ PDF for $\qq$ background are
obtained from a fit to sideband data 
(5.2 GeV/$c^2 <M_{\rm bc}<$5.26 GeV/$c^2$ and $\Delta E>-0.15$ GeV). 
The parameters of the $\Delta t$ PDFs for $\BBbar$ and rare $B$ are obtained 
from a fit to MC.

\begin{figure}
\begin{center}
\includegraphics[width=\columnwidth]{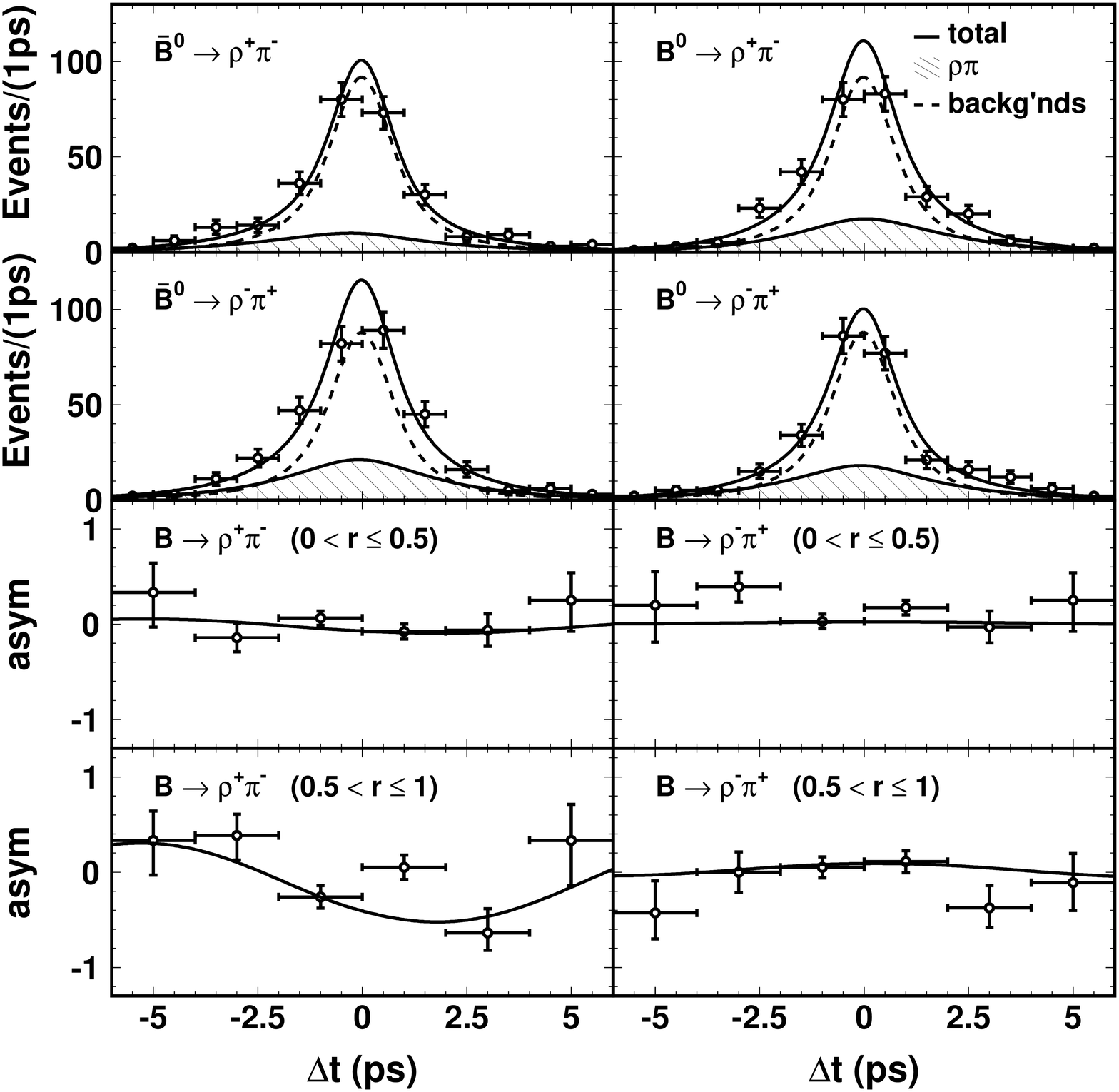}
\caption{\label{fig:acp} $\Delta t$ distributions for $\Bzrp$.
(Top) $B^0$ and $\overline{B}{}^0$ tagged $\rho^+\pi^-$ and 
$\rho^-\pi^+$ candidates.
(Bottom) raw $CP$ asymmetries in high and low $r$ intervals 
for $\rho^+ \pi^-$ and $\rho^- \pi^+$. 
The solid curves show the fit results.
}
\end{center}
\end{figure}

The maximum likelihood fit to the 1,215 $\rho\pi$ candidates gives
$\CrpCP = \Crpvalue \Crpallerr$,
$\dCrpCP = \dCrpvalue \dCrpallerr$,
$\SrpCP = \Srpvalue \Srpallerr$ and
$\dSrpCP = \dSrpvalue \dSrpallerr$, where the first (second) errors are
statistical (systematic). 
The correlation between $\CrpCP$ and $\dCrpCP$ is 0.271 and that
between $\SrpCP$ and $\dSrpCP$ is 0.284, while correlations between
other variables are smaller.
The data and fit result are shown in Fig.~\ref{fig:acp}.

The systematic error in $A_{CP}^{\rho\pi}$ includes a possible background
asymmetry ($\pm0.010$) and charge asymmetry in the tracking ($\pm0.012$).
The charge dependence of tracking efficiency is studied using
$D^0\to K^-\pi^+$ decays from inclusive $D^{*+}\to D^0\pi^+$ and selecting
the momentum region corresponding to $B^0\to \rho^\pm\pi^\mp$ decays.  
The systematic errors for time-dependent measurements include
the uncertainties in the
vertex reconstruction, background fraction, background $\Delta t$ PDF,
wrong-tag fractions, $\rho\pi$ and $\rho{K}$ $\Delta t$ resolution functions, 
physics parameters ($\tau_B$, $\Delta m_d$~\cite{bib:pdg}, 
$\ArKCP$~\cite{bib:babar_rhopi})  
and fitting bias. 
The fitting bias is estimated from MC pseudo-experiments. All other
systematic uncertainties are obtained by varying parameters within their errors
and repeating the fit. 
The dominant source of systematic error is the vertex reconstruction
($^{+0.012}_{-0.055}$ for $\CrpCP$, $^{+0.011}_{-0.038}$ for $\dCrpCP$, 
 $^{+0.094}_{-0.073}$ for $\SrpCP$, 
and $^{+0.089}_{-0.092}$ for $\dSrpCP$).

We perform various consistency checks. 
We examine the stability of the results as the $\mathcal{R}$ selection 
criterion is varied and the asymmetry of the $\Delta t$ distributions for 
events in the sideband region. No significant variation or asymmetry is 
observed. We measure the $B^0$ lifetime with the 
$B^0\to \rho^\pm\pi^\mp$ candidates and find 
$\tau_{B^0} = 1.56 ^{+0.13}_{-0.12}$ ps, which is consistent with the
world average value~\cite{bib:pdg}.  

The extraction of
$\phi_2$ from measurements of time-dependent $CP$ violation parameters in
$B \to \rho^\pm\pi^\mp$ decays has been studied in several
theoretical approaches~\cite{bib:phi2_qcdf,bib:phi2_su3,bib:gronau_zupan}.
A Grossman-Quinn type bound~\cite{bib:grossman_quinn} based on
isospin (SU(2) symmetry) does not
significantly limit the penguin diagram contribution
due to the large branching fraction
for $B^0 \to \rho^0\pi^0$~\cite{bib:belle_rho0pi0}.
Since the number of measurable quantities (six including
${\cal{B}}(B^0\to\rho^+\pi^-)$) are not sufficient to completely describe
the amplitudes for $B^0\to\rho^\pm\pi^\mp$ decay (8 free parameters),
either specific models or additional assumptions
are involved, such as QCD factorization~\cite{bib:phi2_qcdf}
or SU(3) flavor symmetry~\cite{bib:phi2_su3}.
A recent approach assuming broken flavor-SU(3)
implies $\phi _2 = (102 \pm 11 \pm 15)^\circ$
using our results~\cite{bib:gronau_zupan}. The first error is
experimental while the second is the uncertainty due to SU(3) 
breaking effects.

In summary, using $152 \times 10^6$ $B\overline {B}$ pairs, we have measured
$CP$ violation parameters for $\Bzrp$ decays. 
We obtain $\ArpCP = \Arpvalue \Arpstat \Arpsyst$,
$\CrpCP = \Crpvalue \Crpallerr$,
$\dCrpCP = \dCrpvalue \dCrpallerr$,
$\SrpCP = \Srpvalue \Srpallerr$ and
$\dSrpCP = \dSrpvalue \dSrpallerr$.
These give the direct $CP$ violation parameters
$\Apm=\Apmvalue$ $\Apmallerr$
and $\Amp=\Ampvalue$ $\Ampallerr$.  These results are consistent with
a previous measurement~\cite{bib:babar_rhopi}.
We find no significant mixing-induced or direct $CP$ violation in
$\Bzrp$.

We thank the KEKB group for the excellent operation of the
accelerator, the KEK Cryogenics group for the efficient
operation of the solenoid, and the KEK computer group and
the NII for valuable computing and Super-SINET network
support.  We acknowledge support from MEXT and JSPS (Japan);
ARC and DEST (Australia); NSFC (contract No.~10175071,
China); DST (India); the BK21 program of MOEHRD and the CHEP
SRC program of KOSEF (Korea); KBN (contract No.~2P03B 01324,
Poland); MIST (Russia); MESS (Slovenia); NSC and MOE
(Taiwan); and DOE (USA).

\end{document}